\documentclass[12pt, draftclsnofoot, onecolumn]{IEEEtran}
\usepackage{amsmath,bm,amssymb}
\usepackage{cases}

\usepackage{graphicx}
\usepackage[skip=3pt]{caption,subcaption}
\usepackage{epstopdf}

\usepackage{cite}
\usepackage{hyperref}

\usepackage{algorithm,algorithmic}

\newtheorem{proposition}{Proposition}

\newtheorem{lemma}{Lemma}
\usepackage{xspace,colortbl}
\newcommand{\cmmt}[1]{{\color{black}{#1}}}

\begin{document}

\title{Joint Spectrum Reservation and On-demand Request for Mobile Virtual Network Operators}

\author{Yingxiao~Zhang, Suzhi~Bi, and Ying-Jun~Angela~Zhang
	\IEEEcompsocitemizethanks{
		\IEEEcompsocthanksitem
		This work has been presented in part in 2016 IEEE ICCS, Shenzhen, China [18].  
		\IEEEcompsocthanksitem
		Yingxiao~Zhang and Ying-Jun~Angela~Zhang are with the Department of Information Engineering, The Chinese University of Hong Kong, HK. Email:~\{zy012,~yjzhang\}@ie.cuhk.edu.hk.
		\IEEEcompsocthanksitem
		Suzhi~Bi is with the College of Information Engineering, Shenzhen University, Shenzhen,  Guangdong, China. Email:~bsz@szu.edu.cn.}
%\vspace*{-\baselineskip}
%\IEEEcompsocitemizethanks{
%	\IEEEcompsocthanksitem This work was supported in part by General Research Funding (Project number 14209414) from the Research Grants Council of Hong Kong and by the National Basic Research Program (973 program Program number 2013CB336701). 
%	The work of S. Bi is supported in part by the National Natural Science Foundation of China (project no. 61501303) and the Foundation of Shenzhen City (project no. JCYJ20160307153818306).}
}

\maketitle

\begin{abstract}
Wireless network virtualization enables mobile virtual network operators (MVNOs) to develop new services on a low-cost platform by leasing virtual resources from mobile network owners. 
In this paper, we investigate a two-stage spectrum leasing framework, where an MVNO acquires radio spectrum through both advance reservation and on-demand request. 
To maximize its surplus, the MVNO jointly optimizes the amount of spectrum to lease in each stage by taking into account the traffic distribution, random user locations, wireless channel statistics, quality-of-service requirements, and the prices differences.
Meanwhile, the MVNO dynamically allocates the acquired spectrum resources to its mobile subscribers (users) according to fast channel fading in order to maximize the utilization of the resources.
The MVNO's surplus maximization problem is formulated as a tri-level nested optimization problem consisting of dynamic resource allocation (DRA), on-demand request, and advance reservation subproblems.
To solve the problem efficiently, we first analyze the DRA problem, and then use the optimal solution to find the optimal leasing decisions in the two stages.
In particular, we derive a closed-form expression of the optimal on-demand request, and develop a stochastic gradient descent algorithm to find the optimal advance reservation.
For a special case when the proportional fairness utility is adopted, we show that the optimal two-stage leasing scheme is related to the number of users and is irrelevant to user locations.
Simulation results show that the two-stage spectrum leasing scheme can adapt to different levels of traffic and on-demand price variations, and achieve higher surplus than conventional one-stage leasing schemes.
\end{abstract}


\section{Introduction}
Wireless Network Virtualization (WNV) is an emerging technology that provides unprecedented opportunities for mobile virtual network operators (MVNOs) to develop new services by leasing infrastructure and radio resources from mobile network owners (MNOs) \cite{2015:Liang, 2015:Granelli,2015:Bi,2016:Samdanis}.
In contrast to the network function virtualization in upper layers, the virtualization of radio spectrum in the physical layer is a unique problem in WNV, due to the broadcasting and stochastic nature of wireless channels \cite{2016:Richart}.
Unlike other network resources, radio spectrum can be dynamically reused by different links based on the geographic separation between transmission nodes, the transmit powers, the interference cancellation capability, and the quality-of-service requirements \cmmt{\cite{Bagwari1, Bagwari2}}.
As a result, spectrum virtualization is much more complicated and deserves in-depth study.

\cmmt{
Existing approaches for radio spectrum virtualization can be divided into two categories: advance reservation and on-demand request.
In advance reservation, each MVNO reserves a certain amount of spectrum resources from an MNO for a long period of time
\cite{2010:Zaki, 2012:NVS, 2013:Guo, 2014:Kamel}.
In particular, \cite{2010:Zaki} considered four types of contract-based reservations and developed a hypervisor to allocate resources according to the predefined contracts.
Reference \cite{2012:NVS} developed a network substrate to virtualize wireless resources, which enables both bandwidth-based and resource-based reservations.
Reference \cite{2013:Guo} developed a partial resource reservation scheme, where each MVNO reserves a  minimum amount of spectrum and the remaining part is shared among all MVNOs based on their real-time demands.
Reference \cite{2014:Kamel} proposed a dynamic resource allocation scheme that keeps track of both the minimum reservation and the fairness requirement.
In general, advance reservation is convenient for the MNO to pre-plan resource slicing and allocation.
However, due to the uncertainty of traffic realizations and wireless channel fading, under-reservation (over-reservation) may occur when the reserved resource is less (more) than the real-time demands.
In contrast, on-demand request preserves flexibility for MVNOs to order spectrum resources according to the observed traffic demands in real time \cite{2016:Chen,2013:Duan:cognitive,2016:Nguyen,2016:Zhu}.
For example, \cite{2016:Chen} considered virtualization in heterogeneous networks, where an MVNO leases resources from both macro and small cells for a particular realization of users.
Reference \cite{2013:Duan:cognitive} studied cognitive MVNOs, which not only lease resources from the MNO but also access the white space by spectrum sensing.
Reference \cite{2016:Nguyen} developed a real-time trading mechanism under incomplete cost information from the MVNOs.
Reference \cite{2016:Zhu} derives a hierarchical combinatorial auction mechanism, where the leasing decisions vary as fast as the variations of instantaneous wireless channel fading.
The flexibility of on-demand request comes at a high operational cost of frequent calculation and realization of the leasing decisions.
Moreover, there is no guarantee of sufficient spectrum supply in real time.
}

In contrast to conventional one-stage leasing schemes, either advance reservation or on-demand request, this paper proposes a two-stage spectrum leasing scheme, where an MVNO leases spectrum from both advance reservation and on-demand request.
When carefully optimized, the two-stage leasing scheme enjoys the complementary strengths of the two stages.
The first stage, advance reservation, reduces the risk and operation complexity for both MVNOs and MNOs. 
On one hand, advance reservation ensures the MVNOs a baseline amount of spectrum at a relatively low cost. 
On the other hand, it allows the MNOs to pre-plan spectrum slicing and system operation in an early stage. 
Meanwhile, the second stage, on-demand request, preserves flexibility and competition. 
On one hand, the MVNOs can acquire additional spectrum according to the real-time traffic demands, and thus avoid being overly conservative in the first stage. 
On the other hand, the MNO can derive more profit by setting a higher on-demand price according to the real-time competition intensity among multiple MVNOs.
The two-stage leasing framework has been previously considered in cloud networks to serve uncertain demands of computing and storage resources \cite{2014:Chase,2017:Luong}.
However, in WNV, random wireless channel fading introduces a new dimension of uncertainty, rendering it difficult for an MVNO to anticipate its need of spectrum in advance.

\begin{figure}
	\centering	
	\includegraphics[width=0.78\textwidth]{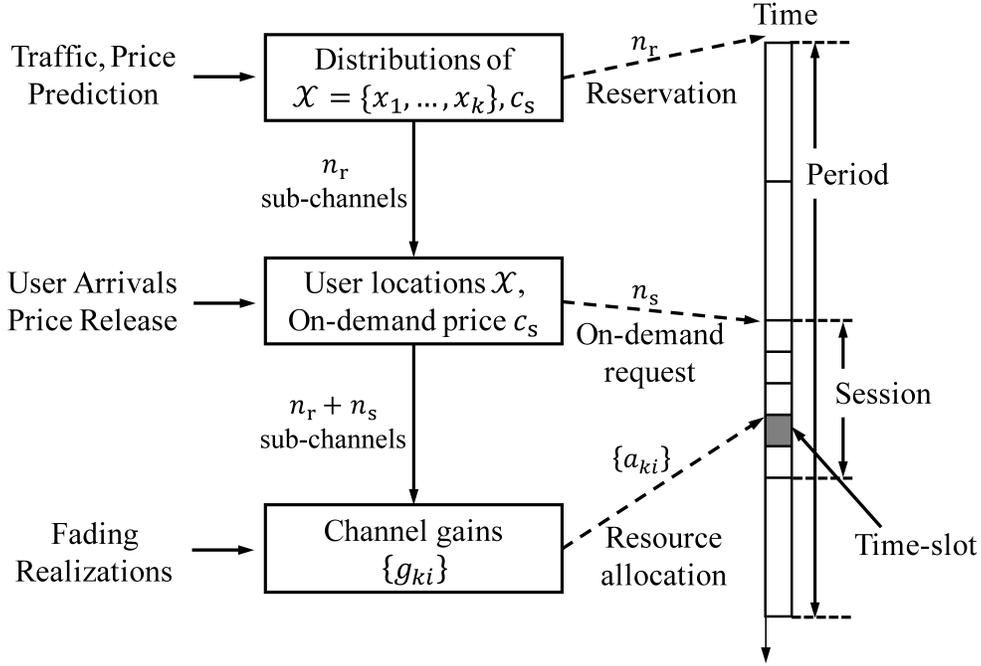}
	\caption{\cmmt{Illustration of the two-stage spectrum leasing framework.}}
	\label{fig:1}
\end{figure}

In this paper, we aim to find the optimal two-stage spectrum leasing scheme for an MVNO.
Specifically, the system operation involves three timescales, as depicted in Fig. \ref{fig:1}.
In the first stage, i.e., advance reservation, the MVNO reserves a certain amount of spectrum resources for a long period of time, which usually covers hours or days.
The decision is optimized according to the statistics of the user traffic and the on-demand price over the period.
In the second stage, i.e., on-demand request, the MVNO decides whether to request additional spectrum after observing the actual number and locations of users it needs to serve.
The on-demand request changes with user arrivals, departures, and movements, and usually varies in a timescale of seconds.
In contrast to \cite{2013:Duan:cognitive,2016:Zhu,2016:Chen}, the on-demand request here does not vary with fast channel fading, and thus involves much lower operational complexity.
Then, in the timescale of milliseconds, the MVNO dynamically allocates the spectrum resources acquired from both advance reservation and on-demand to the users according to the fast channel fading.
To maximize its profit, the MVNO needs to jointly optimize the operations in all three timescales.

The two-stage spectrum leasing problem is naturally formulated as a tri-level nested optimization problem that consists of dynamic resource allocation (DRA), on-demand request, and advance reservation subproblems.
Solving the problem is challenging in two aspects.
First, the nested structure makes the three subproblems closely intertwined.
The decision made in a larger timescale affects the optimization in a smaller timescale.
In turn, the optimal value of a smaller-timescale problem is embedded in the objective function of a large-timescale problem.
Hence, it is critical to analytically characterize the optimal values of the smaller-timescale problems, so that the larger-timescale problems are amenable to efficient solution algorithms.
Secondly, the nested optimization problem is stochastic in the sense that the decision in a larger timescale must be optimized for random network realizations in a smaller timescale.
\cmmt{As a first attempt, we have tried to solve the two-stage spectrum leasing problem in a precedent conference paper \cite{2016:ZhangYX}, which assumes $\alpha$-fair utility functions and constant on-demand price. 
In this paper, we take into account the randomness of the on-demand price and extend the analysis to more general utility functions.}

\cmmt{
In this paper, we address the challenges as follows:
\begin{itemize}
    \item 
    We derive the optimal channel-aware DRA policy under a broad class of utility functions.
	Through rigorous analysis, we characterize the optimal utility as a function of the total number of sub-channels (SCs) acquired from both advance reservation and on-demand request.

	\item 
	Based on the result from DRA, we solve the on-demand request and advance reservation subproblems for general utility functions.
	In particular, we derive a closed-form expression of the optimal number of SCs to request in the on-demand stage.
	Moreover, we develop a stochastic gradient descent (SGD) algorithm to solve the advance reservation problem efficiently. 

	\item 
	For the special case when proportional fairness (PF) utility is adopted, we derive closed-form expressions of the optimal number of SCs to lease during the two stages.
	In other words, the MVNO can find the optimal operations in all three timescales by analytical calculations with negligible computational complexity.
\end{itemize}
Our numerical results show that the derived two-stage leasing scheme can exploit the on-demand price variations to reduce the MVNO's operating cost.
Moreover, it can adapt to different levels of traffic variations and achieve more surplus than conventional one-stage spectrum leasing schemes.
}

\cmmt{
The rest of paper is organized as follows.
In Section \ref{sec:model}, we describe the network model and introduce the two-stage spectrum leasing framework.
We solve the DRA problem for general utility functions in Section \ref{sec:ra}.
Then, we derive the optimal solution of the on-demand request and the advance reservation subproblems in Section \ref{sec:extension}.
Further, in Section \ref{sec:pf}, we analyze the special properties of the two-stage leasing scheme when PF utility is adopted.
The numerical results and discussions are presented in Section \ref{sec:simulation}.
Finally, we conclude this work in Section \ref{sec:conclusion}.
}

\section{System Model and Problem Formulation}
\label{sec:model}
\begin{figure}
	\centering	
	\includegraphics[width=0.42\textwidth]{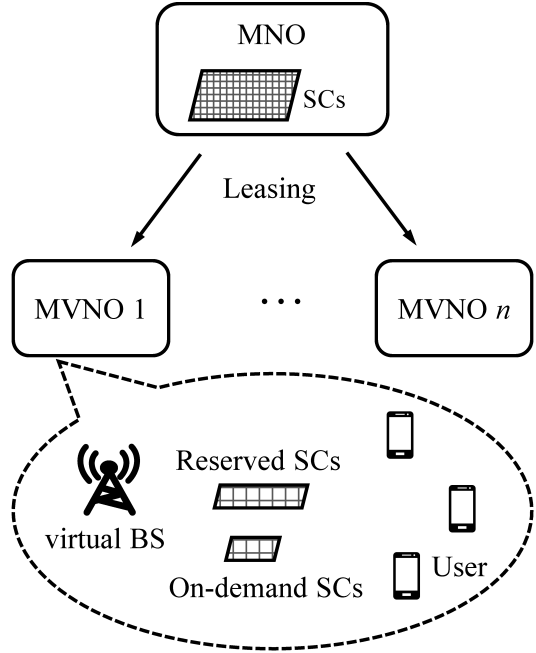}
	\vspace*{\baselineskip}
	\caption{\cmmt{A snapshot of the virtualized mobile network. An MVNO leases sub-channels (SCs) from the MNO by both advance reservation and on-demand request. The acquired SCs are programmed to serve the MVNO's users.}}
	\label{fig:0}
\end{figure}

Consider a single MVNO that acquires radio spectrum from an MNO and programs on the acquired spectrum to serve its users.
In particular, we focus on the downlink transmission in a single OFDMA cell,
	where the spectrum owned by the MNO is divided into a number of sub-channels (SCs).
A snapshot of the virtualized mobile network is shown in Fig. \ref{fig:0}.
Without loss of generality,
	we assume that the cell is a circular area with radius $D$ around a base station (BS) at the origin,
	denoted by $\mathcal{D}(o,D)$.
A set of users requesting services from the MVNO at the same time is denoted by a random point process  $\mathcal{X}=\left\{x_k, \forall k=1,\dots, K\right\}$,
	where each $x_k\in\mathcal{D}$ is the location of the $k$-th user,
	and $K\in\mathbb{Z}_+$ is the total number of users.
We assume that each $x_k$ is uniformly distributed in the cell,
	and the cumulative distribution function (CDF) of $K$ is $F_K(\cdot)$ in the domain $[K_\mathrm{low},K_\mathrm{up}]$,
	where $K_\mathrm{low}$, $K_\mathrm{up}$ are the lower and the upper bounds, respectively.
The user set $\mathcal{X}$ changes with user arrivals, departures, or movements. 
In this way, the traffic statistics is characterized by the distribution of $\mathcal{X}$.
Suppose that transmitted signals are affected by both large-scale path loss and fast Rayleigh fading.
The instantaneous data rate for the $k$-th user on the $i$-th SC is given by
\begin{equation}
b_{ki} = B\log\left(1+\frac{P\ell(\|x_k\|) g_{ki}}{\Gamma N_0 B}\right),
\label{bit}
\end{equation}
where $B$ denotes the bandwidth of each SC,
$P$ denotes the fixed transmit power,
$\ell(\cdot)$ denotes the path-loss function,
$\|\cdot\|$ denotes the Euclidean norm,
$g_{ki}$ denotes the power gain of the i.i.d. Rayleigh fading,
$N_0$ denotes the power spectral density of white noise at the receiver, 
and $\Gamma$ denotes the capacity margin \cite{3gpp:PHY}.

The two-stage spectrum leasing scheme involves operations in three timescales, as depicted in Fig. \ref{fig:1}.
For the sake of clarity, we define \emph{a period, a session, and a time-slot} as three basic time units, during which traffic statistics (i.e., the distribution of $\mathcal{X}$), user locations (i.e., the realization of $\mathcal{X}$), and channel fading (i.e., $\{g_{ki}, \forall k,i\}$) remain unchanged, respectively.
Typically, a period is measured in hours, a session in seconds, and a time-slot in milliseconds.
At the beginning of a period,
	the MVNO reserves a number of SCs for the whole period according to the distribution of $\mathcal{X}$.
Then, at the beginning of each session,
	the MVNO decides whether to request additional SCs for the session according to the observed realization of $\mathcal{X}$.
The acquired SCs from both reservation and on-demand request are dynamically allocated to the users according to fast channel fading in each time-slot.
The MVNO's surplus maximization problem can be formulated as a tri-level nested optimization problem as follows:

\subsubsection{Advance Reservation}
Let $c_\mathrm{r}$ be the reservation price of SCs, and $n_\mathrm{r}$ be the number of reserved SCs.
The problem of finding the optimal reservation is formulated as
\begin{equation}
\underset{n_\mathrm{r}\in \mathbb{Z}_{+}}{\text{maximize}} \ 
-c_\mathrm{r} n_\mathrm{r}+\mathbf{E}_{\mathcal{X},c_\mathrm{s}} \left[Q\left(\mathcal{X},c_\mathrm{s},n_\mathrm{r}\right)\right],
\label{op:long}
\end{equation}
where the first term in the objective function is the reservation cost, and the second term is the expected surplus from each session.
The expectation is taken over all possible user realizations $\mathcal{X}$ and the on-demand price $c_\mathrm{s}$ in a session.
$Q\left(\mathcal{X}, c_\mathrm{s}, n_\mathrm{r}\right)$ denotes the surplus from a session given $\mathcal{X}$, $c_\mathrm{s}$ and $n_\mathrm{r}$, and will be defined explicitly later.
We assume that the reserved SCs cannot be released back to the MNO until the end of the period.

The advance reservation should take into account the traffic uncertainty from the user side, as well as the price uncertainty from the MNO side.
Specifically, the user set $\mathcal{X}$ varies across sessions due to random user arrivals, departures, and mobility.
Moreover, the on-demand price $c_\mathrm{s}$ varies according to the competition level among MVNOs in each session.
We assume that the MVNO can estimate the distribution of $c_\mathrm{s}$ during a period according to historical information.
\cmmt{Let $F_{c_\mathrm{s}}(\cdot)$ and $f_{c_\mathrm{s}}(\cdot)$ denote the cumulative distribution function (CDF) and probability density function (PDF) of $c_\mathrm{s}$, respectively.
	Note that constant $c_\mathrm{s}$ as has been studied in \cite{2016:ZhangYX} is a special case where the variance of $c_\mathrm{s}$ is zero.}
Typically, the MNO sets $c_\mathrm{r}$ lower than $c_\mathrm{s}$	in order to encourage reservation in advance.
We will discuss the impact of the price differences in details through rigorous analysis in the following sessions.

\subsubsection{On-demand Request}
At the beginning of each session, the MVNO decides whether to request additional SCs 
	according to the observed user set $\mathcal{X}$ and the on-demand price $c_\mathrm{s}$ announced by the MNO.
Let $n_\mathrm{s}$ be the number of SCs to request in the session.
The problem of finding the optimal on-demand request is formulated as
\begin{equation}
Q\left(\mathcal{X},c_\mathrm{s}, n_\mathrm{r}\right) = \ \underset{n_\mathrm{s}\in \mathbb{Z}_{+}}{\text{maximize}}\ -c_\mathrm{s} n_\mathrm{s} + u_\mathrm{g} G\left(\mathcal{X},n_\mathrm{r}+n_\mathrm{s}\right),
\label{op:short}
\end{equation}
where $u_\mathrm{g}$ is a fixed scaler that converts the utility into monetary unit, 
	and $G\left(\mathcal{X},n\right)$ denotes the maximum utility achieved by serving the user set $\mathcal{X}$ with $n$ SCs.
The first term in the objective function of (\ref{op:short}) is the cost of the on-demand SCs, and the second term is the utility-based income of serving the users.
The optimal value $Q\left(\mathcal{X},c_\mathrm{s}, n_\mathrm{r}\right)$ is referred to as the surplus from each session, which becomes part of the MVNO's total  profit.
Note that unlike the number of SCs from the on-demand request changes across sessions.

\subsubsection{Dynamic Resource Allocation}
The maximum system utility $G(\mathcal{X},n)$ of a session is achieved by physical layer DRA.
In particular, the MVNO dynamically allocates $n=n_\mathrm{r}+n_\mathrm{s}$ SCs to the $K$ users in $\mathcal{X}$ according to the fast channel fading in each time-slot.
Benefiting from the independent channel variations across users and SCs, DRA can substantially improve the system utility due to multiuser diversity.
Let $U(\bar{r})$ denote the utility function of each user, where $\bar{r}$ is the average throughput over the session.
We assume $U(\bar{r})$ as a continuously differentiable increasing concave function which captures the satisfaction of the user when received throughput $\bar{r}$ \cite{2005:Song}.
By deploying different utility functions, the MVNO can balance the trade-off between throughput maximization and fairness among users.

Let $g_{ki}[t]$ denote the fast channel fading between the BS and the $k$-th user on the $i$-th SC at the $t$-th time-slot, 
	and $b_{ki}[t]$ denote the corresponding achievable data rate calculated by (\ref{bit}).
Let $a_{ki}[t]\in[0,1]$ denote the fraction of airtime of the $i$-th SC allocated to the $k$-th user in the $t$-th time-slot,
	and $A=\{a_{ki}[t],\forall k, i, t\}$ denote all the allocation decisions in the session.
Then, the average throughput of the $k$-th user (for $k=1,\dots,K$) is given by
\begin{equation}
\bar{r}_k=\frac{1}{T}
\sum_{t=1}^{T} \sum_{i=1}^n  a_{ki}[t] b_{ki}[t],
\label{thoughput}
\end{equation}
where $T$ is the total number of time-slots in a session.

The problem of finding the optimal SC allocation for all time-slots can be formulated as
\begin{subequations}
	\begin{align}
	G\left(\mathcal{X},n \right) = \ \underset{A}{\text{maximize}}  \ & \sum_{k=1}^{K} U(\bar{r}_k)
	\\
	\text{subject to} \ 
	&\sum_{k=1}^{K} a_{ki}[t] = 1, \ \forall i,\, t \label{3b}
	\\
	& a_{ki}[t] \geq 0,\ \forall k,\,i,\,t.  \label{3c}
	\end{align}
	\label{op:ra}
\end{subequations}

In particular, $G(\emptyset,n)=0$, meaning that no utility is 0 if the user set is empty.
Note that $G(\mathcal{X},n)$ may be negative if $U(\bar{r})$ represents the penalties.
For example, when $U(\bar{r})=-1/\bar{r}$, the DRA problem in (\ref{op:ra}) tries to minimize the overall transmission delay.

From the description above, 
	we can see that the three optimization problems are nested.
The optimal values of the problems in smaller timescales, i.e., $Q(\mathcal{X},c_\mathrm{s},n_\mathrm{r} )$ and $G(\mathcal{X},n)$, 
	are embedded in the objective functions in larger timescales (\ref{op:long}) and (\ref{op:short}).
In turn, the decisions in larger timescales $n_\mathrm{r}$ and $n_\mathrm{s}$ are parameters of the problems in smaller timescales.

In what follows, we first study the DRA problem in (\ref{op:ra}) in Section \ref{sec:ra}.
Using the results from DRA, we then derive the optimal reservation and on-demand request for general utility functions in Section \ref{sec:pf}.
Further, we analyze the special properties of the two-stage leasing scheme when PF utility function is adopted in Section \ref{sec:extension}.

\section{Dynamic Resource allocation}
\label{sec:ra}
We first study the DRA problem in (\ref{op:ra}),
 which tries to maximize the utilization of the acquired SCs according to fast channel fading in each time-slot.
 
\subsection{Optimal DRA Policy}
Let $\Lambda=\{\lambda_i[t],\forall i, t\}$ and $\mathcal{V}=\{\nu_{ki}[t],\forall i,j,t\}$ denote the Lagrangian multipliers with respect to constraints (\ref{3b}) and (\ref{3c}), respectively.
The Lagrange function is given by
\begin{equation}
\begin{aligned}
L\left(A,\Lambda,\mathcal{V}\right)	
&= \sum_{k=1}^{K} U\left(\bar{r}_k \right)	
+ \sum_{i=1}^n \sum_{t=1}^{T} \lambda_i[t] \left(1- \sum_{k=1}^{K} a_{ki}[t] \right)
+  \sum_{i=1}^n \sum_{k=1}^{K} \sum_{t=1}^{T}\nu_{ki}[t] a_{ki}[t].
\end{aligned}
\end{equation}
Let $A^*$ denote the optimal solution and $\bar{\bm{r}}^*=[\bar{r}_1^*,\dots,\bar{r}_K^*]$ denote the corresponding optimal average throughputs for all users. 
The following KKT conditions hold:
\begin{subequations}
\begin{align}
&\frac{\partial L}{\partial a_{ki}[t]}\bigg|_{A^*} =  \nabla U(\bar{r}_k^*) \frac{b_{ki}[t]}{T} - \lambda_i[t] + \nu_{ki}[t] = 0, \; \forall \, k,i,t
\label{eqn:lag_a}\\
&\sum_{k=1}^{K} a^*_{ki}[t] = 1, \; \forall i,t 
\label{eqn:lag_b}\\
& \ a^*_{ki}[t] \ge 0,\ \nu_{ki}[t] \ge 0,\ \nu_{ki}[t] a^*_{ki}[t] = 0, \; \forall k,i,t.
\label{eqn:lag_c}
\end{align}
\end{subequations}
Here, $\nabla U(r_0)=\frac{d U(\bar{r})}{d\bar{r}} \big|_{\bar{r}=r_0}$ is the first-order derivative of $U(\bar{r})$ evaluated at $\bar{r}=r_0$.

We can infer from (\ref{eqn:lag_a}) and (\ref{eqn:lag_c}) that if $a^*_{ki}[t] > 0$, then $\nu_{ki}[t] = 0$ and
\begin{equation}
\nabla U(\bar{r}_k^*) b_{ki}[t] = T \lambda_i[t] \ge \nabla U(\bar{r}_{k'}^*) b_{k'i}[t], \ \forall k'\neq k.
\end{equation}
In other words, the $i$-th SC is allocated to the user (users) that has (have) the largest $\nabla U(\bar{r}_k^*) b_{ki}[t]$. As $b_{ki}[t]$ is drawn from a continuous distribution,	the probability that two or more users have the same value of $\nabla U(\bar{r}_k^*) b_{ki}[t]$ is zero. Therefore, each SC is exclusively allocated to a single user in each time-slot.
The optimal allocation policy is described by:
\begin{equation}
a_{ki}^*[t]= \begin{cases}
1, & k = \arg \max_{k'} \nabla U(\bar{r}_{k'}^*) b_{k'i}[t] \\
0, & \text{otherwise}.
\end{cases}
\label{policy}
\end{equation}

The policy in (\ref{policy}) states that the SC in each time-slot is allocated to the user which can obtain a relatively high bit rate with respect to a function of its average throughput.
Notice that the SC allocation policy in (\ref{policy}) requires the knowledge of the optimal throughputs $\bar{\bm{r}}^*$, which will be derived in the next subsection.

\subsection{Optimal Average Throughput}
With the SC allocation policy in (\ref{policy}), the average throughput of each user is given by
\begin{equation}
\bar{r}_k^* = \frac{1}{T} \sum_{t=1}^{T} \sum_{i=1}^n b_{ki}[t]
\mathbf{1}_{\left\{\nabla U(\bar{r}_k^*) b_{ki}[t] > \nabla U_{k'}(\bar{r}_{k'}^*) b_{k'i}[t],
	\forall k'\neq k \right\}},
\label{ra:1}
\end{equation}
$\forall k=1,\dots, K$, where $\mathbf{1}_{\{\cdot\}}$ is the indicator function with value $1$ if the argument is true, and $0$ otherwise. 
By the assumption that $T$ is sufficiently large so that channel fading process is ergodic, the time-averaged throughput in (\ref{ra:1}) can be transfered to the expectation with respect to fast fading,
	which is given by
\begin{equation}
\begin{aligned}
	\bar{r}_k^*& = \sum_{i=1}^n \mathbf{E}_{\{b_{ki}\}} \left[ b_{ki}
	\mathbf{1}_{\left\{\nabla U(\bar{r}_k^*) b_{ki} > \nabla U_{k'}(\bar{r}_{k'}^*) b_{k'i},\forall k'\neq k \right\}}
	\right]
	\\
	& = \sum_{i=1}^n \mathbf{E}_{b_{ki}}\left[b_{ki} 
	\mathbf{Pr}\left(b_{k'i}< \frac{\nabla U (\bar{r}_k^*)}{\nabla U(\bar{r}_{k'}^*)} b_{ki}, \forall k'\neq k  \bigg| b_{ki} \right)	\right]
	\\
	& = \sum_{i=1}^n \mathbf{E}_{b_{ki}}\left[b_{ki} \prod_{k'\neq k}
	\mathbf{Pr}\left(b_{k'i}< \frac{\nabla U (\bar{r}_k^*)}{\nabla U(\bar{r}_{k'}^*)} b_{ki} \right) \right],
\end{aligned}
\label{ra:2}
\end{equation}
where the last step follows the independence of channel fading across users.
Note that the division by $\nabla U(r)$ requires $\nabla U(r)>0$, which is true for continuously increasing utility functions.
Since fast fadings are independently and identically distributed,
	$b_{ki}$ for all $i$ follow a same distribution,
	and the CDF and PDF, denoted by $F_{b_k}(\cdot)$ and $f_{b_k}(\cdot)$ respectively, can be calculated from the distribution of Rayleigh fading according (\ref{bit}).
Note that $F_{b_k}(\cdot)$ and $f_{b_k}(\cdot)$ only depend on the user's location $x_k$.
Then, (\ref{ra:2}) can be written as
\begin{equation}
\begin{aligned}
	\bar{r}_k^*
	& = n \mathbf{E}_{b_k}\left[b_k \prod_{k'\neq k} F_{b_{k'} } \left(
	\frac{\nabla U (\bar{r}_k^*)}{\nabla U(\bar{r}_{k'}^*)} b_k\right) \right]
	\\
	& = n \int_{0}^\infty \eta \prod_{k'\neq k} F_{b_{k'} } \left(
	\frac{\nabla U (\bar{r}_k^*)}{\nabla U(\bar{r}_{k'}^*)} \eta \right) f_{b_k}(\eta)\, d\eta.
\end{aligned}
\label{ra:3}
\end{equation}

From (\ref{ra:3}), we can see that the optimal throughput $\bar{\bm{r}}^*$ is the root of the following nonlinear equation system
\begin{equation}
r_k = n \Phi_k(\bm{r}), \; \forall k=1,\dots,K,
\label{ra:5}
\end{equation}
where
\begin{equation}
\Phi_k(\bm{r}) = \int_{0}^\infty \eta \prod_{k'\neq k} F_{b_{k'} } \left( \frac{\nabla U (r_k)}{\nabla U(r_{k'})} \eta\right) f_{b_k}(\eta)\, d\eta.
\label{ra:4}
\end{equation}

By Brouwer's fixed-point theorem \cite{1976:fixedpoint}, there is at lease one root for the system in (\ref{ra:5}).
From (\ref{ra:4}), we can see that $\Phi_k(\bm{r})$ is a decreasing function of $r_k$, since $\nabla U(r_k)$ is a decreasing function of $r_k$.
This means that there is at most one $r_k$ satisfying $r_k = n \Phi_k(\bm{r})$ given that all $\{\bar{r}_{k'}, \forall k'\neq k \}$ are fixed.
Therefore, the equation system in (\ref{ra:5}) has a unique root.

The average throughputs of all users can be calculated numerically by solving the non-linear system (\ref{ra:5}).
Substituting the optimal average throughputs $\bar{\bm{r}}^*$ into (\ref{policy}), we can obtain the optimal DRA policy for each set of users that maximizes the system total utility with $n$ SCs.
Note that the DRA policy is computed at the beginning of each session according to the distribution of fast fading.
Then, at the beginning of each time-slot, the SCs are allocated to serve users according to the estimated fast channel fading.

\subsection{Optimal System Utility}
With the optimal DRA policy, we now derive the analytical expression of the optimal value $G(\mathcal{X},n)$,
	which is required to solve the on-demand request problem in (\ref{op:short}).
With a bit abuse of notation, we use $\bar{\bm{r}}^*(n)$ to denote the optimal throughput vector achieved with $n$ SCs.
By solving the equation system (\ref{ra:5}),
	we can obtain $\bar{\bm{r}}^*(n)$ and the corresponding optimal utility $G(\mathcal{X},n)=\sum_{k=1}^{K} U(\bar{r}_k^*(n))$.
In general, there is no closed-form expression of $G(\mathcal{X},n)$.
To preserve analytical tractability, we focus on a broad class of utility functions
	where $\bar{r}^*(n)$ increases linearly with $n$.

\begin{lemma}
	The average throughput of each user achieved by the optimal DRA policy in (\ref{policy}) increases linearly with the number of SCs, i.e.,
	\begin{equation}
	\bar{r}_k^*(n) = n \bar{r}_k^*(1), \text{ for } k=1,\dots, K,
	\label{ra:7}
	\end{equation}
	if the utility function satisfies the condition
	\begin{equation}
	\frac{\nabla U(r_1)}{\nabla U(r_2)} = \frac{\nabla U(r_1/n)}{\nabla U(r_2/n)},
	\label{condition}
	\end{equation}
	for any non-negative rates $r_1,\,r_2>0$ and positive integer $n=1,2,\dots$.
	\label{lmm}
\end{lemma}

\begin{IEEEproof}
	With the property in (\ref{condition}), the average rate in (\ref{ra:3}) can be written as
	\begin{equation}
	\begin{aligned}
	\bar{r}^*_k(n)/n =\Phi_k(\bar{\bm{r}}^*(n)) =\Phi_k(\bar{\bm{r}}^*(n)/n),
	\text{ for } k=1,\dots, K.
	\end{aligned} 
	\label{ra:6}
	\end{equation}
	This means that $\bar{\bm{r}}^*(n)/n$ for all $n$ are roots of the equation system $r_k = \Phi_k(\bm{r}) $ for $k=1,\dots, K$.
	As the system has a unique root $\bar{\bm{r}}^*(1)$, we have
	\begin{equation}
	\bar{\bm{r}}^*(n)/n = \bar{\bm{r}}^*(1), \ \forall n=1,\dots,
	\end{equation}
	which leads to the result in (\ref{ra:7}).
\end{IEEEproof}

The most well-known class of utility functions that satisfies condition (\ref{condition}) is the $\alpha$-fair utility functions, which are defined by \cite{2011:Wang:policy}
\begin{equation}
	\begin{aligned}
	U(\bar{r}) = \begin{cases}
	\frac{1}{1-\alpha}\bar{r}^{(1-\alpha)}, & \text{if } \alpha\ge0, \alpha\neq 1,
	\\
	\log(\bar{r}), &\text{if } \alpha=1,
	\end{cases}
	\end{aligned}
\label{ex:11}
\end{equation}
where $\alpha$ is the degree of fairness and $\bar{r}$ is the average throughput of a user.
Specifically, DRA based on $\alpha$-fair utility turns out to be a throughput maximization problem when $\alpha=0$,
	and becomes a delay minimization problem when $\alpha=2$.
Moreover, when $\alpha=1$, proportional fairness is achieved among the users with the logarithm utility function.
Note that not all concave increasing utility functions satisfy condition (\ref{condition}).
For example, the exponential utility function where $U(\bar{r})=1-e^{-\bar{r}}$ and the positive diminishing return where $U(\bar{r})=\ln(1+\bar{r})$ \cite{2013:Huang}.

When (\ref{condition}) holds, the average data rate is a linear function of the number of SCs.
As a result, we only need to solve the nonlinear system (\ref{ra:5}) once for $\bar{\bm{r}}^*(1)$, and then the average throughputs  with a general $n$ are obtained accordingly.
Note that $\bar{\bm{r}}^*(1)$ is uniquely determined by the user locations $\mathcal{X}$ and the distribution of fast channel fading.
Moreover, the optimal utility can be calculated by
\begin{equation}
	G(\mathcal{X},n)=\sum_{k=1}^{K} U(\bar{r}_k^*(n))=\sum_{k=1}^{K} U(n\bar{r}^*_k(1)),
	\label{ra:9}
\end{equation}
which will be used to derive the optimal advance reservation and on-demand request in the next section.

\section{Optimal Two-Stage Leasing Scheme}
\label{sec:extension}
\cmmt{
	In this section, we investigate the optimal on-demand request and advance reservation for the utility functions that satisfy the condition in Lemma \ref{lmm}. 
}

\subsection{Optimal On-demand Request}
\label{sec:ex:short}
With the optimal utility from DRA in (\ref{ra:9}), the on-demand request problem in (\ref{op:short}) becomes
\begin{equation}
	Q(\mathcal{X},c_\mathrm{s}, n_\mathrm{r}) = \max_{n_\mathrm{s}\in \mathbb{Z}_+} -c_\mathrm{s} n_\mathrm{s} + u_\mathrm{g} \sum_{k=1}^{K} U\left((n_\mathrm{r}+n_\mathrm{s})\bar{r}^*_k(1)\right).
	\label{ex:2}
\end{equation}
By the property in (\ref{condition}), we can compute the first-order derivative of the objective function with respect to $n_\mathrm{s}$ as
\begin{equation}
	\begin{aligned}
		& -c_\mathrm{s} + u_\mathrm{g}\sum_{k=1}^{K} \bar{r}^*_k(1)\nabla U\left((n_\mathrm{r}+n_\mathrm{s})\bar{r}^*_k(1)\right)
		\\
		=& -c_\mathrm{s} + u_\mathrm{g}\nabla U(n_\mathrm{r}+n_\mathrm{s})\sum_{k=1}^{K} \bar{r}^*_k(1) \frac{\nabla U\left((n_\mathrm{r}+n_\mathrm{s})\bar{r}^*_k(1)\right)}{\nabla U(n_\mathrm{r}+n_\mathrm{s})}
		\\
		=& -c_\mathrm{s} + u_\mathrm{g} \nabla U(n_\mathrm{r}+n_\mathrm{s}) \sum_{k=1}^{K} \bar{r}^*_k(1) \frac{\nabla U\left(\bar{r}^*_k(1)\right)}{\nabla U(1)}
		\\
		=&-c_\mathrm{s} + u_\mathrm{g} \nabla U(n_\mathrm{r}+n_\mathrm{s}) \Theta(\mathcal{X}) ,
	\end{aligned}
	\label{ex:1}
\end{equation}
where
\begin{equation}
	\Theta(\mathcal{X}) = \frac{1}{\nabla U(1)} \sum_{k=1}^{K} \bar{r}^*_k(1) \nabla U\left(\bar{r}^*_k(1)\right).
	\label{ex:3}
\end{equation}
Note that $\Theta(\mathcal{X})$ is uniquely determined by the user locations $\mathcal{X}$, and the value can be calculated by solving the non-linear system (\ref{ra:5}) for $n=1$.
Since the objective in (\ref{ex:2}) is a concave function of $n_\mathrm{s}$, setting the first-order derivative (\ref{ex:1}) to zero yields the optimal solution
\begin{equation}
	n_\mathrm{s}^* = \max \left(\nabla U^{-1}\left(\frac{c_\mathrm{s}}{u_\mathrm{g} \Theta(\mathcal{X})} \right)-n_\mathrm{r}, 0 \right),
	\label{ex:sln}
\end{equation}
where $\nabla U^{-1}(\cdot)$ is the inverse function of $\nabla U(\cdot)$.
The integer solution can be obtained by rounding $n_\mathrm{s}^*$.
From (\ref{ex:sln}), we can see that the MVNO uses the reserved $n_\mathrm{r}$ SCs to serve a baseline amount of traffic,
	and requests additional SCs when $c_\mathrm{s} < u_\mathrm{g} \Theta(\mathcal{X}) \nabla U(n_\mathrm{r})$. 
Due to the concavity of $U(\cdot)$, $\nabla U(\cdot)$ is a decreasing function, 
and hence $n_\mathrm{s}^*$ decreases with a higher $c_s$ or a smaller $\Theta(\mathcal{X})$.
This means that the MVNO requests less SCs when observing a higher on-demand price or a user realization with a smaller utility metric $\Theta(\mathcal{X})$.

A special case is the linear utility function where $U(\bar{r}) = \bar{r}$.
In this case, $n_\mathrm{s}^*$ is either zero or infinity, depending on $\Theta(\mathcal{X})=\sum_{k=1}^K \bar{r}^*_k(1)$.
This is because that both the leasing cost and the users' utility increase linearly with $n_\mathrm{s}$, and the slope depends on $\Theta(\mathcal{X})$.
In practice, the number of SCs for sale is usually limited.
Hence, the optimal strategy for the MVNO is to lease as much SCs as possible if $c_\mathrm{s} < u_\mathrm{g} \Theta(\mathcal{X}) $, or no on-demand request otherwise.

With the optimal on-demand request (\ref{ex:sln}),	we can compute the corresponding optimal surplus in the session as
\begin{equation}
	\begin{aligned}
		& Q(\mathcal{X},c_\mathrm{s},n_\mathrm{r}) =
		\begin{cases}
			u_\mathrm{g} \sum_{k=1}^{K} U(n_\mathrm{r} \bar{r}^*_k(1)),
			\hfill \text{if } c_\mathrm{s} > u_\mathrm{g} \Theta(\mathcal{X}) \nabla U(n_\mathrm{r}) ;\,&
			\\ 
			c_\mathrm{s} n_\mathrm{r} - c_\mathrm{s}\nabla U^{-1}\left(\frac{c_\mathrm{s}}{u_\mathrm{g} \Theta(\mathcal{X})} \right) 
			+ u_\mathrm{g} \sum_{k=1}^K U(\nabla U^{-1}\left(\frac{c_\mathrm{s}}{u_\mathrm{g} \Theta(\mathcal{X})} \right) \bar{r}^*_k(1)),
			\quad \text{ otherwise.} &
		\end{cases}
	\end{aligned}
	\label{ex:4}
\end{equation}
Specially, $Q(\emptyset, c_\mathrm{s}, n_\mathrm{r})=0$, corresponding to zero payoff of an idle session.
Here, we use the real-value solution $n_\mathrm{s}^*$ as an approximation to the integer solution in calculating $Q(\mathcal{X}, c_\mathrm{s}, n_\mathrm{r})$.
The approximation error in calculating the optimal advance reservation in the next subsection is negligible, since the error can be averaged out in $\mathbf{E}[Q(\mathcal{X}, c_\mathrm{s}, n_\mathrm{r})]$.

\subsection{Optimal Advance Reservation}
\label{sec:ex:long}
With the expression of $Q(\mathcal{X},c_\mathrm{s},n_\mathrm{r})$ in (\ref{ex:4}),
	we now solve the advance reservation problem in (\ref{op:long}).
Let $J(n_\mathrm{r})$ and $\nabla J(n_\mathrm{r})$ denote the objective function of (\ref{op:long}) and its first-order derivative, respectively.
With (\ref{condition}), we can compute $\nabla J(n_\mathrm{r})$ as
\begin{equation}
	\begin{aligned}
		\nabla J(n_\mathrm{r}) &= -c_\mathrm{r}+\mathbf{E}_{\mathcal{X},c_\mathrm{s}}
		\left[\min \left(c_\mathrm{s}, u_\mathrm{g} \Theta(\mathcal{X}) \nabla U(n_\mathrm{r}) \right)\right].
	\end{aligned}
	\label{ex:5}
\end{equation}

\begin{proposition}
	When $\mathbf{E}[c_\mathrm{s}] \le c_\mathrm{r}$, the MVNO makes no reservation, i.e., $n_\mathrm{r}^*=0$.
	In other words, all SCs are acquired from on-demand request in each session.
	\label{lmm:2}
\end{proposition}
\begin{IEEEproof}
	From (\ref{ex:5}), we have
	\begin{equation}
	\begin{aligned}
	\nabla J(n_\mathrm{r}) 	&=-c_\mathrm{r}+\mathbf{E}_{\mathcal{X},c_\mathrm{s}}
	\left[\min \left(c_\mathrm{s}, u_\mathrm{g} \Theta(\mathcal{X}) \nabla U(n_\mathrm{r}) \right)\right]
	\\
	&\le -c_\mathrm{r} + \mathbf{E}[c_\mathrm{s}],
	\end{aligned}
	\end{equation}
	due to the fact that $c_s \ge \min \left(c_\mathrm{s}, u_\mathrm{g} \Theta(\mathcal{X}) \nabla U(n_\mathrm{r}) \right)$.
	When $\mathbf{E}[c_\mathrm{s}] \le c_\mathrm{r}$, $\nabla J(n_\mathrm{r})\le 0$,
		meaning that $J(n_\mathrm{r})$ is a decreasing function of $n_\mathrm{r}$.
	Therefore, the optimal non-negative solution is $n_\mathrm{r}^*= 0$, which completes the proof.
\end{IEEEproof}

Proposition \ref{lmm:2} shows that a price discount is essential to motivate the MVNO to place reservation in advance.
From the MNO's perspective, advance reservation has advantages of risk-free incomes and simple operation.
Hence, the MNO usually sets a discount to encourage MVNOs to reserve resources for a long period of time.

For the case $\mathbf{E}[c_\mathrm{s}] > c_\mathrm{r}$, as $J(n_\mathrm{r})$ is a concave function, the optimal real-value solution $n_\mathrm{r}^*$ satisfies the first-order condition
\begin{equation}
	\begin{aligned}
	0 = -c_\mathrm{r} + \mathbf{E}_{\mathcal{X},c_\mathrm{s}}
	\left[\min \left(c_\mathrm{s}, u_\mathrm{g} \Theta(\mathcal{X}) \nabla U(n_\mathrm{r}^*) \right)\right].
	\end{aligned}
	\label{ex:sln_long}
\end{equation}
From (\ref{ex:sln_long}), we can see that $n_\mathrm{r}^*$ increases as $c_\mathrm{r}$ decreases or $c_\mathrm{s}$ increases.
This matches the intuition that the MVNO reserves more SCs in advance for a lower reservation price or to avoid more expensive on-demand request.

Due to the lack of closed-form expression of $\Theta(\mathcal{X})$,
neither $J(n_\mathrm{r})$ nor $\nabla J(n_\mathrm{r})$ can be computed analytically
even with the distribution functions of $c_\mathrm{s}$ and $\mathcal{X}$.
Therefore, there is no analytical expression for the optimal reservation $n_\mathrm{r}^*$.
As $J(n_r)$ is concave, we develop a stochastic gradient descent (SGD) algorithm, as presented in Algorithm \ref{alg1} in the next page, to find the optimal reservation numerically.
The key idea is to approximate the real gradient by that of a sampled user set in each iteration.
Similar to (\ref{ex:5}), we can compute the approximated gradient of a sample $\mathcal{X}$ by
\begin{equation}
\begin{aligned}
\Delta(\mathcal{X}) &= -c_\mathrm{r} + \mathbf{E}_{c_\mathrm{s}}
\left[\min \left(c_\mathrm{s}, u_\mathrm{g} \Theta(\mathcal{X}) \nabla U(n_\mathrm{r}) \right)\right]
\\
&= -c_\mathrm{r} + \int_0^{u_\mathrm{g} \Theta(\mathcal{X}) \nabla U(n_\mathrm{r})} \left(1-F_{c_\mathrm{s}}(\eta)\right)d\eta,
\end{aligned}
\label{ex:6}
\end{equation}
where the last step follows that
\begin{equation}
\begin{aligned}
\mathbf{E}_{c_\mathrm{s}} \left[\min \left(c_\mathrm{s}, c\right) \right] 
&= \int_0^{c} \eta dF_{c_\mathrm{s}}(\eta)+c\left(1-F_{c_\mathrm{s}}(c)\right)
\\
&= c F_{c_\mathrm{s}}(c)-\int_0^c F_{c_\mathrm{s}}(\eta)d\eta +c\left(1-F_{c_\mathrm{s}}(c)\right)
\\
&=\int_0^c \left(1-F_{c_\mathrm{s}}(\eta)\right)d\eta.
\end{aligned}
\label{ex:7}
\end{equation}
With the CDF of $c_\mathrm{s}$, the approximate gradient can be computed efficiently by (\ref{ex:6}).
By \cite[Theorem 3.4]{2016:SGD}, the SGD algorithm returns an $\epsilon$-approximate solution after $L=O(1/\epsilon^2)$ iterations, when the step size is set to $\eta[l]=1/\sqrt{l}$ for the $l$-th iteration step.
The integer solution can be obtained by rounding the optimal real-value solution.

\begin{algorithm*}
	\caption{SGD for finding the optimal advance reservation}
	\label{alg1}
	\begin{algorithmic}[1]
		\REQUIRE initial value $n_\mathrm{r}[1] \in \mathbb{Z}_+$, step size $\{\eta[l],1\le l\le L\}$
		\FOR{$l=1$ \TO $L-1$}
		\STATE Sample a set of users and their locations $\mathcal{X}=\{x_1,\dots, x_K\}$
		\STATE Calculate the average throughputs $\bar{\bm{r}}^*(1)$ by solving (\ref{ra:5}) for $n=1$
		\STATE Calculate $\Theta(\mathcal{X})$ by (\ref{ex:3})
		\STATE Calculate the gradient of the sample $\Delta(\mathcal{X})$ by (\ref{ex:6})
		\STATE Update and project :
		\begin{eqnarray}
		\begin{aligned}
		n_\mathrm{r}[l+1] &= n_\mathrm{r}[l] + \eta[l] \Delta(\mathcal{X}) \\
		n_\mathrm{r}[l+1] &= \max (n_\mathrm{r}[l+1], 0)
		\end{aligned}
		\end{eqnarray}
		\ENDFOR
		\ENSURE $n_\mathrm{r}^* = \frac{1}{L} \sum_{l=1}^L n_\mathrm{r}[l]$
	\end{algorithmic}
\end{algorithm*}

\cmmt{
\section{Application to Proportional Fairness Utility}
\label{sec:pf}
In this section, we apply the analytical results in the previous section to PF utility,
		and analyze the unique properties of the two-stage leasing scheme that are not reflected in that for general utility functions.
Formally, the PF utility function is defined as \cite{2008:Angela:PF}}
\begin{equation}
U(\bar{r}) = \log(\bar{r}),
\label{eqn:pf_def}
\end{equation}
for $\bar{r}\ge 0$, which is a special case of $\alpha$-fair utility functions for $\alpha=1$.
The first-order derivative is given by $\nabla U(\bar{r})=1/\bar{r}$,
	which satisfies the condition in Lemma \ref{lmm}.

\subsection{Optimal On-demand Request}
\label{sec:pf_short}
Substituting the first-order derivative $\nabla U(\bar{r})=1/\bar{r}$ into (\ref{ex:3}), we get $\Theta(\mathcal{X})=K$.
Then, the optimal on-demand request in (\ref{ex:sln}) can be simplified as
\begin{equation}
n_\mathrm{s}^* = \max \left(\frac{u_\mathrm{g}}{c_\mathrm{s}} K -n_\mathrm{r}, 0\right).
\label{short:sln}
\end{equation}

In contract to the general case (\ref{ex:sln}), the optimal on-demand request for PF in (\ref{short:sln}) is only related to the number of users in the session and is irrelevant to the user locations.
This is because that in the sum utility $G(\mathcal{X},n)=K \log(n) + \sum_{k=1}^{K} \log\left( \bar{r}^*_k(1)\right)$, the information about user locations $\mathcal{X}$ only appears in the second additional term that is irrelevant to leasing decision $n$.
The physical meaning of (\ref{short:sln}) can be interpreted as follows.
The MVNO needs no additional SCs if there are only a small number of active users or the on-demand price is very high, i.e., $u_\mathrm{g}K \le n_\mathrm{r} c_\mathrm{s}$.
Otherwise, the MVNO requests additional SCs, whose number increases linearly with the number of users.
Moreover, we can see that the MVNO's on-demand request is proportional to the inverse of $c_\mathrm{s}$,
which is the same as the widely used demand curve in telecommunication systems \cite{2013:Huang}.

\subsection{Optimal Advance Reservation}
\label{sec:pf_long}
Substituting $\nabla U(\bar{r})=1/\bar{r}$ and $\Theta(\mathcal{X})=K$ into (\ref{ex:sln_long}),
	we have
\begin{equation}
\begin{aligned}
0 &=-c_\mathrm{r} + \mathbf{E}_{K,c_\mathrm{s}}\left[\min \left(c_\mathrm{s}, u_\mathrm{g} K /n_\mathrm{r}^* \right) \right]
\end{aligned}
\label{long:1}
\end{equation}
As shown in Proposition \ref{lmm:2}, $n_\mathrm{r}^*>0$ for $\mathbf{E}[c_\mathrm{s}] > c_\mathrm{r}$.
In this case, (\ref{long:1}) can be rewritten as
\begin{equation}
\begin{aligned}
 c_\mathrm{r} n_\mathrm{r}^* &= \mathbf{E}_{K,c_\mathrm{s}}\left[\min \left(c_\mathrm{s}n_\mathrm{r}^*, u_\mathrm{g} K\right) \right]
 \\
 &= \int_0^{u_\mathrm{g} K_\mathrm{up}} \mathbf{Pr}\left(\min \left(c_\mathrm{s}n_\mathrm{r}^*, u_\mathrm{g} K\right) \ge \eta \right) d\eta
 \\
 &= \int_0^{u_\mathrm{g} K_\mathrm{up}} \left( 1- F_{c_\mathrm{s}}(\eta/n_\mathrm{r}^*) \right) 
 	\left( 1- F_{K}(\eta/u_\mathrm{g}) \right) d\eta,
\end{aligned}
\label{long:2}
\end{equation}
where the first step follows that $\mathbf{E}[Y]=\int \mathbf{Pr}(Y\ge \eta) d\eta$ for a non-negative random variable $Y$.
With the CDFs of $c_\mathrm{s}$ and $K$, we can obtain $n_\mathrm{r}^*$ by solving the equation in (\ref{long:2}).
Note that $n_\mathrm{r}^*$ in the case of PF only depends on the distribution of $K$ instead of the distribution of user locations $\mathcal{X}$ as in general utility functions.
In other words, the MVNO can easily compute the optimal number of SCs to lease in the advance reservation stage without applying Algorithm \ref{alg1}.

\begin{proposition}
In the case of PF utility function, the average cost of the optimal two-stage leasing scheme increases linearly with the average number of users in each session. That is
	\begin{equation}
	c_\mathrm{r} n_\mathrm{r}^*+\mathbf{E}[c_\mathrm{s} n_\mathrm{s}^*] = u_\mathrm{g}\mathbf{E}[K].
	\end{equation}
	\label{lmm:3}
\end{proposition}
\begin{IEEEproof}
	From (\ref{long:1}), we can compute the cost on the reserved SCs by
	\begin{equation}
	\begin{aligned}
	c_\mathrm{r} n_\mathrm{r}^* &= \mathbf{E}_{K,c_\mathrm{s}}\left[\min \left(c_\mathrm{s}, u_\mathrm{g} K /n_\mathrm{r}^* \right) \right] \cdot n_\mathrm{r}^*
	\\
	&=\mathbf{E}_{K,c_\mathrm{s}}\left[\min(c_\mathrm{s} n_\mathrm{r}^*, u_\mathrm{g} K )\right].
	\end{aligned}
	\label{lmm3:1}
	\end{equation}
	Further, from (\ref{short:sln}), we can calculate the average cost on the on-demand request by
	\begin{equation}
	\begin{aligned}
	\mathbf{E}[c_\mathrm{s} n_\mathrm{s}^*] &= \mathbf{E}_{K,c_\mathrm{s}} \left[
	c_\mathrm{s} \cdot \max \left(\frac{u_\mathrm{g}}{c_\mathrm{s}} K -n_\mathrm{r}^*, 0\right)
	\right]
	\\
	&=\mathbf{E}_{K,c_\mathrm{s}} \left[
	\max \left(u_\mathrm{g} K , c_\mathrm{s} n_\mathrm{r}^*\right) -c_\mathrm{s} n_\mathrm{r}^*
	\right].
	\end{aligned}
	\label{lmm3:2}
	\end{equation}
	Adding (\ref{lmm3:1}) to (\ref{lmm3:2}), we can obtain the total average cost as
	\begin{equation}
	\begin{aligned}
	c_\mathrm{r} n_\mathrm{r}^*+\mathbf{E}[c_\mathrm{s} n_\mathrm{s}^*]
	&=\mathbf{E}_{K,c_\mathrm{s}}\left[\min(c_\mathrm{s} n_\mathrm{r}^*, u_\mathrm{g} K )+\max(c_\mathrm{s} n_\mathrm{r}^*, u_\mathrm{g} K)-c_\mathrm{s} n_\mathrm{r}^*\right]
	\\
	&=\mathbf{E}_{K,c_\mathrm{s}} \left[c_\mathrm{s} n_\mathrm{r}^*+ u_\mathrm{g} K -c_\mathrm{s} n_\mathrm{r}^*\right]
	\\
	&=u_\mathrm{g}\mathbf{E}[K],
	\end{aligned}
	\end{equation}
	which leads to the proof.
\end{IEEEproof}

Proposition \ref{lmm:3} shows that more investment is needed for a busy period with a large number of users.
However, the allocation of the investment into advance reservation and on-demand request depends on the variation of $K$ and $c_\mathrm{s}$, which will be discussed with numerical results in the next section.

\section{Numerical Results and Discussion}
\label{sec:simulation}
In this section, we evaluate numerically the performance of the derived two-stage leasing scheme.
The users of a tagged MVNO are randomly dropped inside a circular cell with radius $D=1$ km.
The number of users in each session $K$ is a random integer uniformly distributed in the interval $[0,16]$.
The wireless channel is Rayleigh faded.
A standard path-loss model with path-loss exponent $3.67$ is adopted \cite{3gpp:PHY}.
The average received SNR at the cell edge is $-6$ dB.
Suppose that the reservation price $c_\mathrm{r}$ is normalized to $1$ and $u_\mathrm{g}=5$.
The on-demand price is uniformly distributed in $[0.8, 1.8]$, unless otherwise specified.
We consider general $\alpha$-fair utility functions with $\alpha=1$ and $\alpha=0.8$ as two examples.
Notice that $\alpha=1$ corresponds to PF and the results are given in Section \ref{sec:pf}.
%For the case with $\alpha=0.8$, we use Algorithm \ref{alg1} to calculate $n_\mathrm{r}^*$,	which converges within $10^6$ iterations.
Extensive simulations show that the use of other distributions or utility functions does not change the conclusions in this paper, and thus are omitted for brevity. 
\cmmt{We run system-level simulations on MATLAB software.}

\cmmt{
\begin{figure*}
	\centering
	\includegraphics[width=0.67\textwidth]{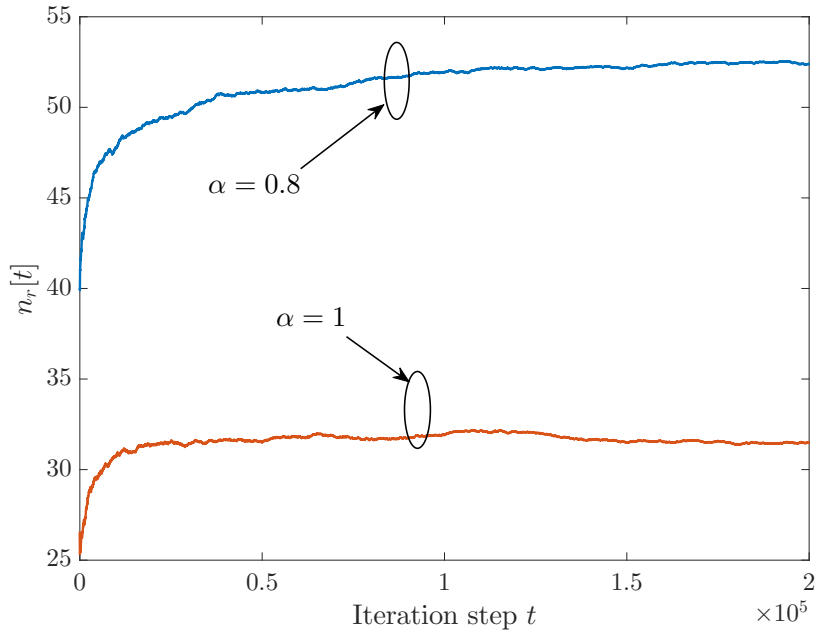}
	\caption{Convergence process of the SGD algorithm to find the optimal reservation.}
	\label{fig:sgd}
\end{figure*}

Fig. \ref{fig:sgd} shows the convergence process of Algorithm \ref{alg1}.
The solution $n_\mathrm{r}[l]$ is plotted as a function of the step $l$.
The main complexity of Algorithm \ref{alg1} is to sample enough realizations of user set, which can be obtained from historical records.
Since the sampling is performed offline, the complexity of Algorithm \ref{alg1} is not an issue for practical implementations.
With the optimal advance reservation, the MVNO can calculate the optimal on-demand request in each session by the closed-form expression in (\ref{ex:sln}).
The overhead to acquire user locations and the on-demand price at the beginning of each session is very low compared with the duration of the session.
The leased SCs are allocated to users following the optimal DRA policy in (9),
	where channel estimation has been standardized in physical layer protocols \cite{3gpp:PHY}.
With our analytical results, the MVNO can implement two-stage leases and DRA with low operational complexity.
}

\begin{figure*}
	\centering
	\begin{subfigure}[b]{0.6\textwidth}
		\subcaption{}
		\includegraphics[width=\textwidth]{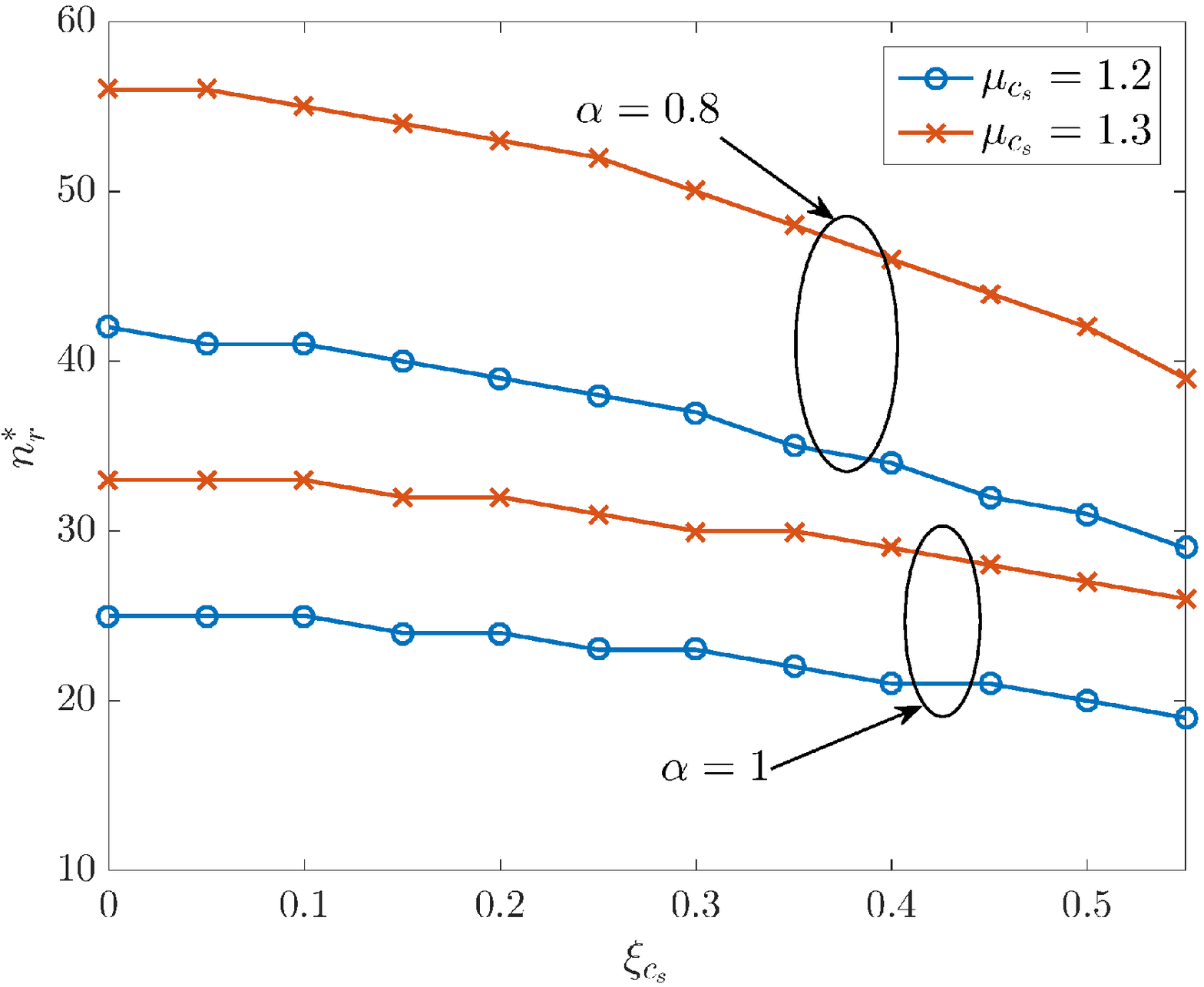}
		\label{fig:2a}
	\end{subfigure}
	\\
	\begin{subfigure}[b]{0.6\textwidth}
		\subcaption{}
		\includegraphics[width=\textwidth]{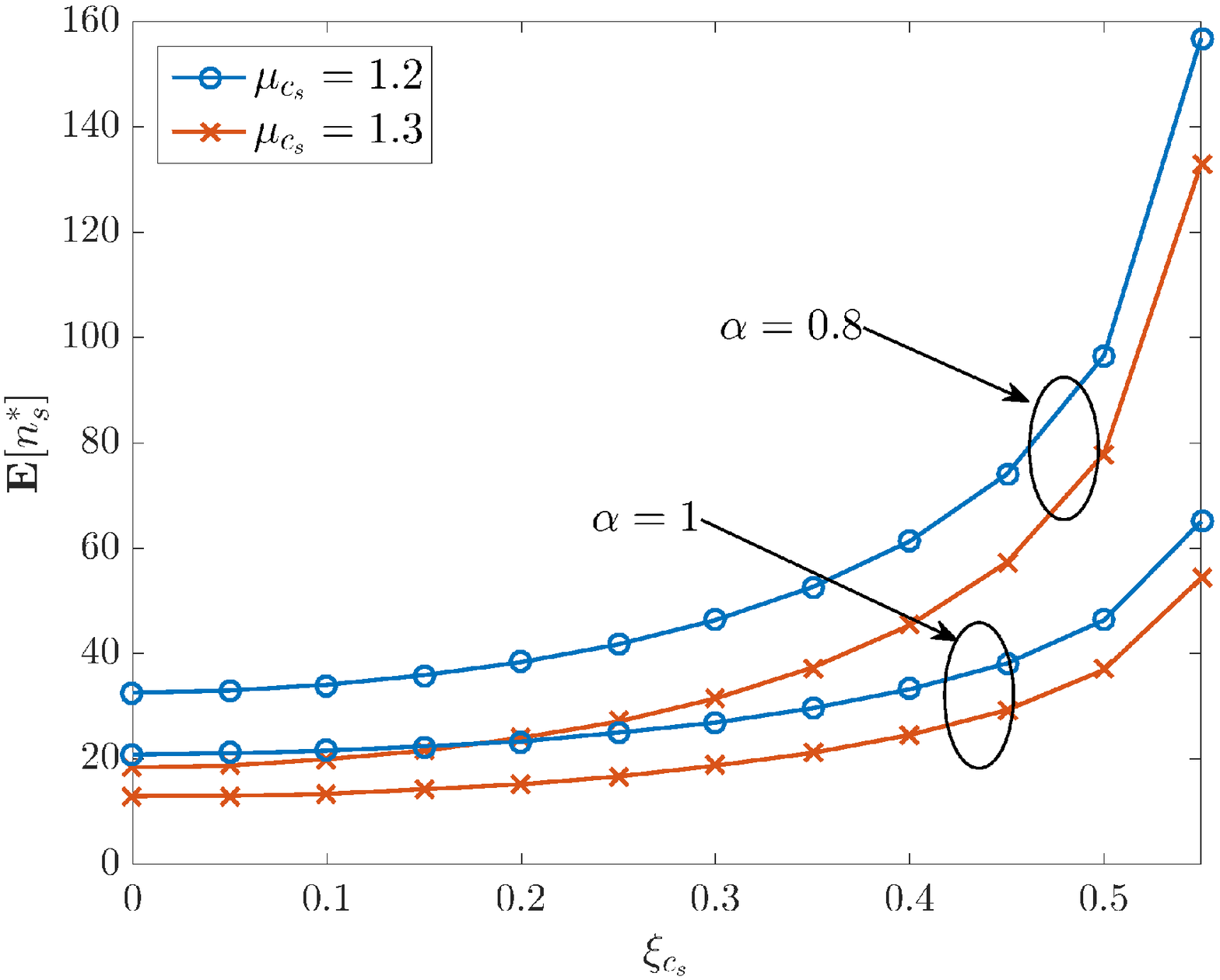}
		\label{fig:2b}
	\end{subfigure}
	\caption{Optimal two-stage leasing decisions versus on-demand price variation $\xi_{c_\mathrm{s}}$.
		(a) Optimal advance reservation $n_\mathrm{r}^*$.
		(b) Expectation of the optimal on-demand request $\mathbf{E}\left[n_\mathrm{s}^*\right]$.
		%The mean of the on-demand price $\mu_{c_\mathrm{s}}$ equals 1.2 or 1.3. 
		%The utility fairness factor $\alpha$ equals $0.8$ or $1$.
	}
	\label{fig:2}
\end{figure*}

As we have shown in Section \ref{sec:extension}, the optimal  leasing decisions in the two stages depend on the variations of on-demand price $c_\mathrm{s}$ and user set $\mathcal{X}$. 
In this section, we will investigate the impact of the variations of $c_\mathrm{s}$ and $\mathcal{X}$ through numerical simulations.
In particular, the variation level of a random variable $z$ is measure by the coefficient of variation, which is defined as
\begin{equation}
\xi_z = \frac {\sigma_z}{\mu_z},
\label{sim:1}
\end{equation}
where $\sigma_z$ and $\mu_z$ are the standard deviation and the mean of the random variable $z$, respectively.
With $\xi_z$, we can compare the variation levels of $z$ under different means.

\subsection{Impact of On-demand Price Variation}
In Fig. \ref{fig:2},
	we fix the mean of the on-demand price $\mu_{c_\mathrm{s}}$, and vary the coefficient of variation $\xi_{c_\mathrm{s}}$ to investigate the effect of the on-demand price variation on the optimal two-stage leasing scheme.
For uniform distribution, the support of $c_\mathrm{s}$ is given by $\left[(1- \sqrt{3} \xi_{c_\mathrm{s}})\mu_{c_\mathrm{s}}, (1+ \sqrt{3} \xi_{c_\mathrm{s}})\mu_{c_\mathrm{s}} \right]$.
\cmmt{In particular, $\xi_{c_\mathrm{s}}=0$ corresponds to the case of constant $c_\mathrm{s}$ over the entire period.}
We focus on the case $\mu_{c_\mathrm{s}}> c_\mathrm{r}$ to avoid the trivial solution $n_\mathrm{r}^*=0$ as shown in Proposition \ref{lmm:2}.
Specifically, $\mu_{c_\mathrm{s}}$ is set to be $1.2$ and $1.3$ in the figures.

In Fig. \ref{fig:2a} and \ref{fig:2b},
	we plot the optimal advance reservation $n_\mathrm{r}^*$ and the expectation of the optimal on-demand request $\mathbf{E}[n_\mathrm{s}^*]$ as functions of $\xi_{c_\mathrm{s}}$, respectively. 
We can see that as $\xi_{c_\mathrm{s}}$ becomes larger, $n_\mathrm{r}^*$ decreases and $\mathbf{E}[n_\mathrm{s}^*]$ increases,
	meaning that the MVNO places less reservation in advance and makes more on-demand request
	when the on-demand price has a larger variation.
Moreover, this trend holds for utility functions with different fairness factors $\alpha$ and for different $\mu_{c_\mathrm{s}}$.
For a larger $\mu_{c_\mathrm{s}}$, more SCs are reserved in advance and less SCs are leased on demand due to the higher cost of on-demand request. 

\begin{figure*}
	\centering
	\begin{subfigure}[b]{0.6\textwidth}
		\subcaption{}
		\includegraphics[width=\textwidth]{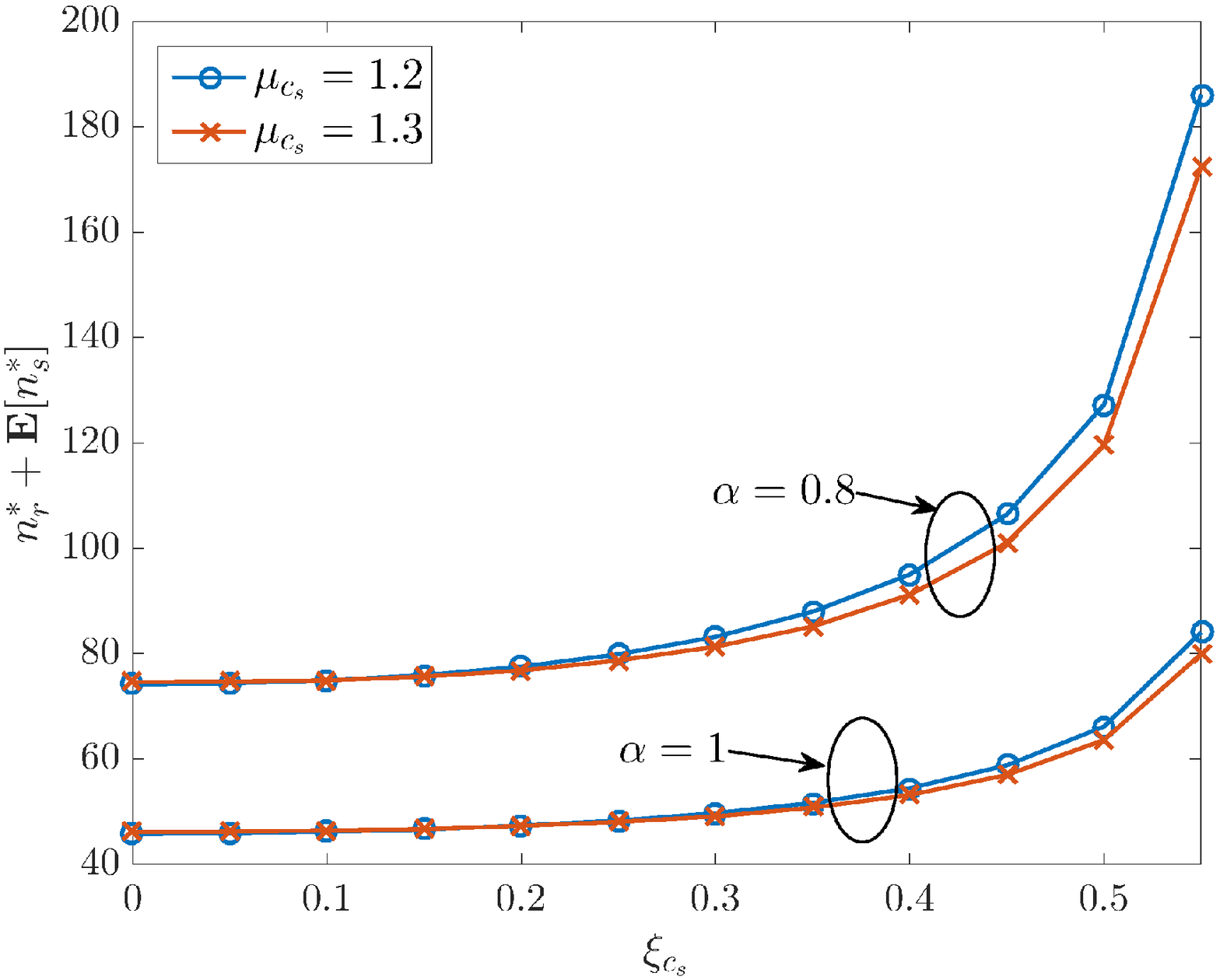}
		\label{fig:2c}
	\end{subfigure}
	\begin{subfigure}[b]{0.6\textwidth}
		\subcaption{}
		\includegraphics[width=\textwidth]{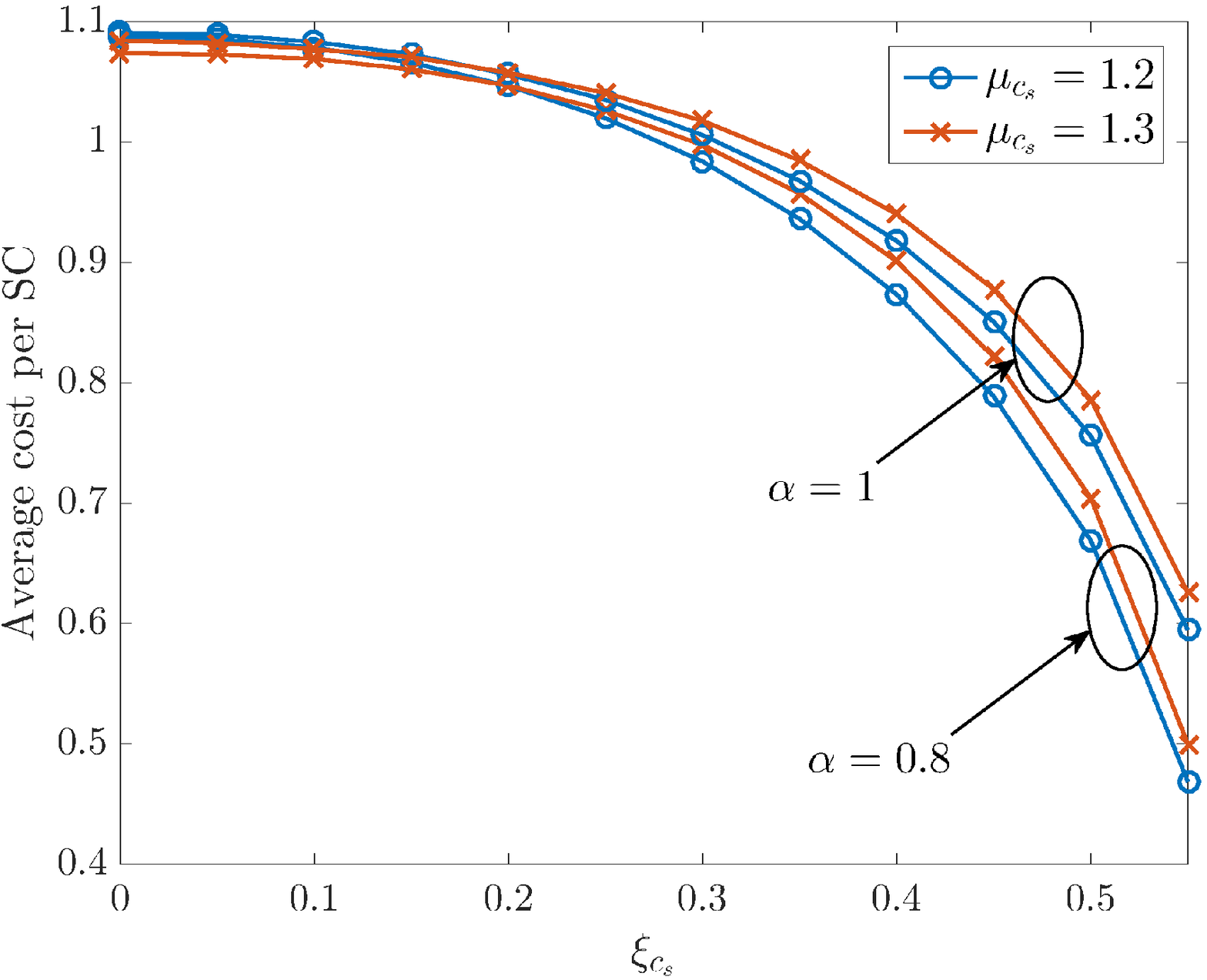}
		\label{fig:2d}
	\end{subfigure}
	\caption{Total leasing amount and average cost versus on-demand price variation $\xi_{c_\mathrm{s}}$.
		(a) Expected total number of SCs per session $n_\mathrm{r}^*+\mathbf{E}[n_\mathrm{s}^*]$.
		(b) Average cost per SC $\frac{c_\mathrm{r} n_\mathrm{r}^*+\mathbf{E}[c_\mathrm{s} n_\mathrm{s}^*]}{n_\mathrm{r}^*+\mathbf{E}[n_\mathrm{s}^*]}$. 
		%The mean of the on-demand price $\mu_{c_\mathrm{s}}$ equals 1.2 or 1.3. 
		%The utility fairness factor $\alpha$ equals $0.8$ or $1$.
	}
	\label{fig:25}
\end{figure*}

We further show the average total number of SCs used for each session and the average leasing cost per SCs in Fig. \ref{fig:2c} and \ref{fig:2d}, respectively.
From Fig. \ref{fig:2c}, we can see that more SCs are leased in total as $\xi_{c_\mathrm{s}}$ becomes larger.
This is because that the lower percentile of $c_\mathrm{s}$ becomes lower when the variance is large. 
Sometimes, the realizations of $c_\mathrm{s}$ may be even lower than $c_\mathrm{r}$,
	which provides MVNO the opportunity to lease on-demand SCs at a low cost. 
Indeed, as shown in Fig. \ref{fig:2d}, the average leasing cost per SCs becomes lower when $\xi_{c_\mathrm{s}}$ increases.
This implies that the derived two-stage leasing scheme can exploit the short-term low on-demand price to lower MVNO's cost.
As a result, more SCs are ordered in total for a larger $\xi_{c_\mathrm{s}}$.
In addition, we can see that the total number of leased SCs increases and the average cost decreases for a lower $\mu_{c_\mathrm{s}}$.
This matches the intuition that the MVNO consumes more SCs for lower on-demand price.

\subsection{Impact of Traffic Variation}
\begin{figure*}
	\centering
	\begin{subfigure}[b]{0.6\textwidth}
		\subcaption{}
		\includegraphics[width=\textwidth]{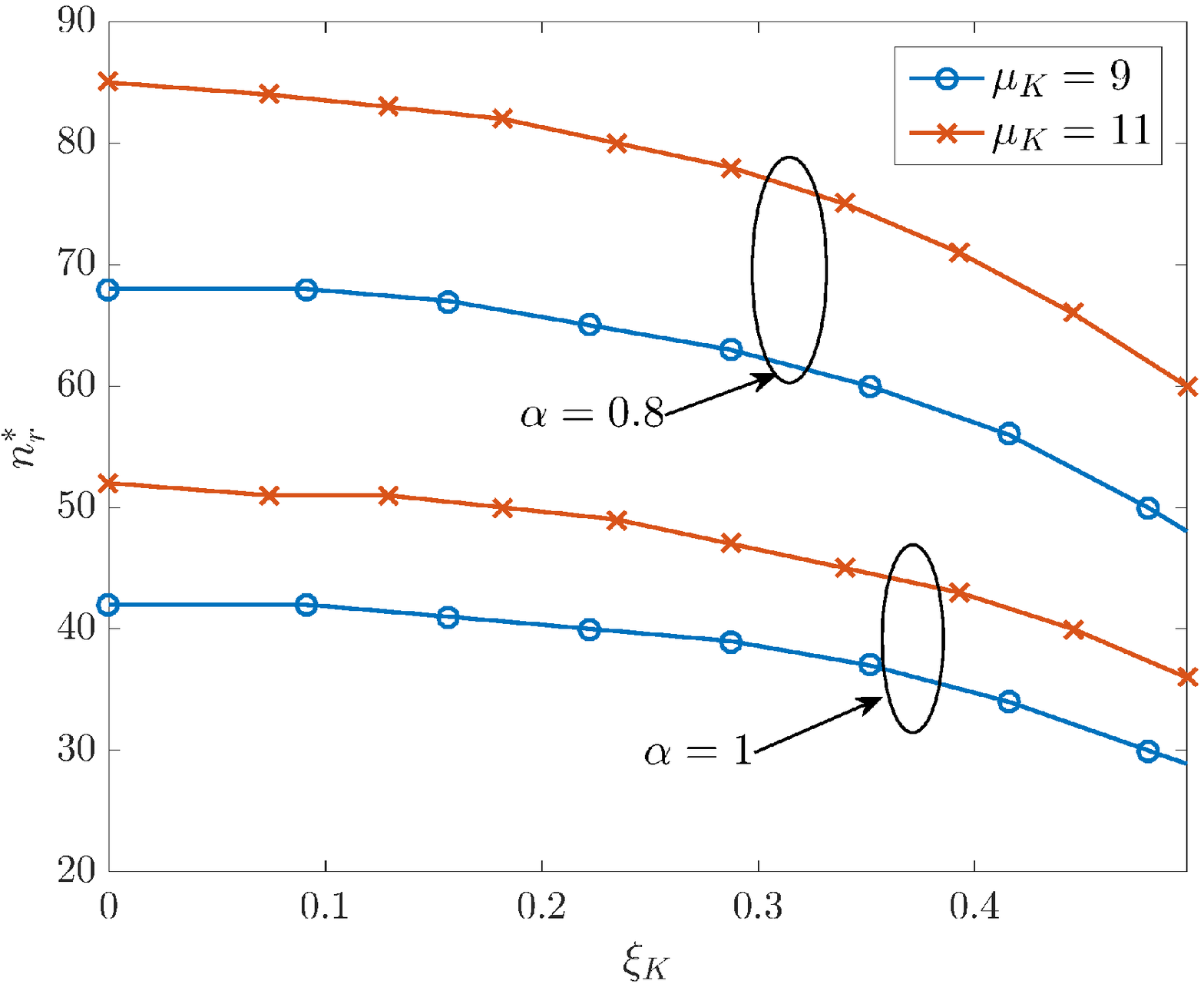}
		\label{fig:3a}
	\end{subfigure}
	~
	\begin{subfigure}[b]{0.6\textwidth}
		\subcaption{}
		\includegraphics[width=\textwidth]{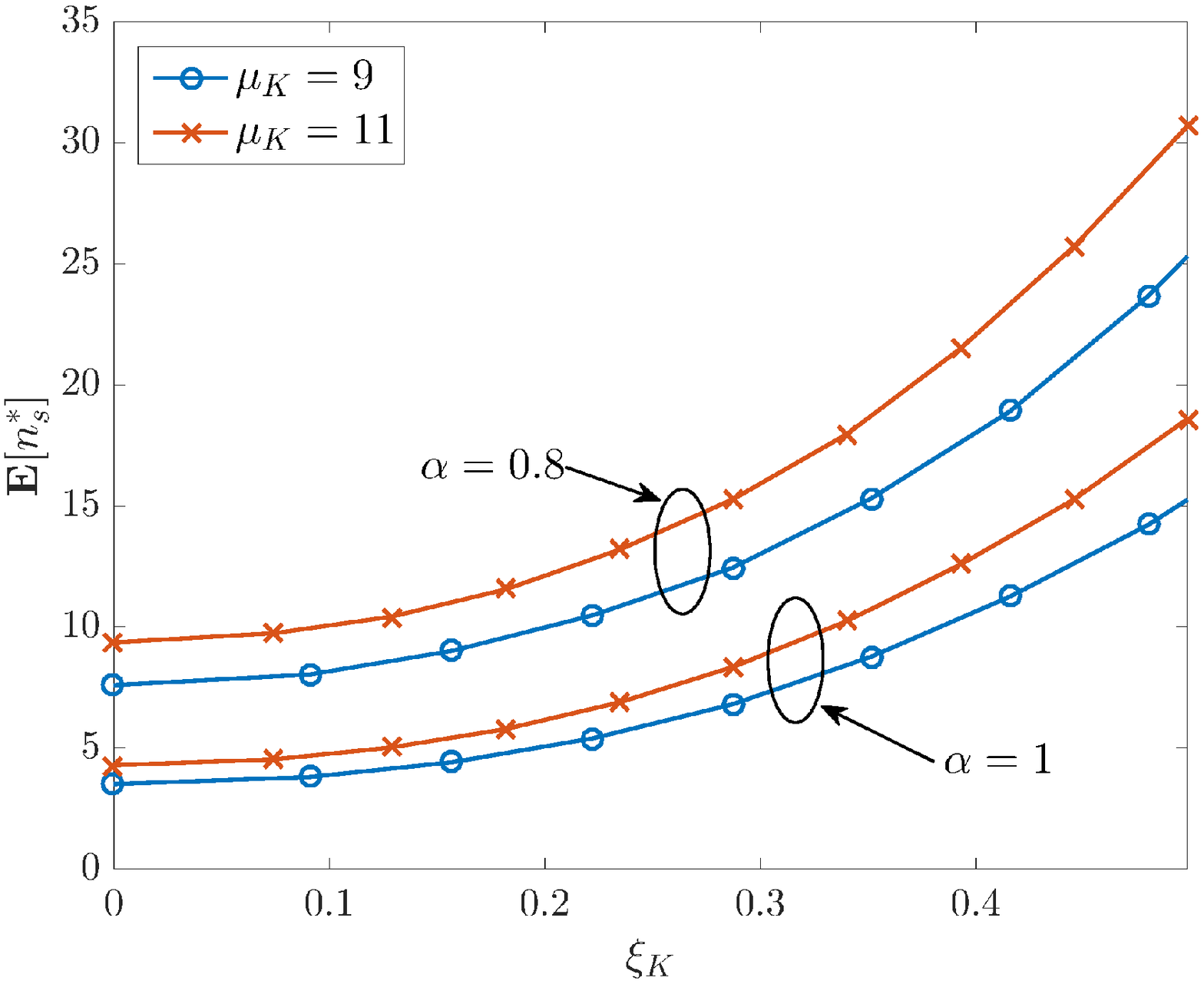}
		\label{fig:3b}
	\end{subfigure}

	\caption{Optimal two-stage leasing decisions versus traffic intensity variation $\xi_{K}$. 
		(a) Optimal advance reservation $n_\mathrm{r}^*$.
		(b) Expectation of the optimal on-demand request $\mathbf{E}\left[n_\mathrm{s}^*\right]$.
		%The average number of users $\mu_{K}$ equals 9 or 11. 
		%The utility fairness factor $\alpha$ equals $0.8$ or $1$.
	}
	\label{fig:3}
\end{figure*}

One advantage of the two-stage leasing scheme is that the MVNO has the flexibility of adapting its leasing request according to the real-time traffic realizations during the on-demand request phase.
In Fig. \ref{fig:3}, we fix the mean of traffic intensity $\mu_{K}$ and vary the coefficient of variation $\xi_{K}$
	to investigate the impact of traffic variation on the optimal leasing decisions. 
From Fig. \ref{fig:3a} and \ref{fig:3b}, we can see that when $\xi_{K}$ increases,
	the optimal reservation $n_\mathrm{r}^*$ decreases, and the expectation of the optimal on-demand request $\mathbf{E}[n_\mathrm{s}^*]$ increases.
This is intuitive in the sense that the MVNO tends to rely more on on-demand request while less on advance reservation to deal with more uncertain traffic demands. 
Moreover, we can see that this observation applies to different traffic intensities $\mu_K$ and different fairness factors $\alpha$.
It can be seen that more SCs are purchased in both reservation and on-demand stages for a larger $\mu_{K}$, 	
	which matches our intuition that the MVNO needs more SCs to serve more intense traffic.

\subsection{\cmmt{Comparison with One-stage Leasing Schemes}}
For comparison, we consider one-stage spectrum leasing schemes, reservation-only and on-demand-only, as benchmarks.
In the reservation-only scheme as used in \cite{2013:Guo}, the MVNO prescribes a number of SCs at the beginning of a period and has no other chances to purchase spectrum throughout the period.
The reservation problem can be formulated similarly to the stochastic programming in (\ref{op:long}), except that $\mathbf{E}_{\mathcal{X},c_\mathrm{s}} \left[Q\left(\mathcal{X},c_\mathrm{s},n_\mathrm{r}\right)\right]$ is replaced by $\mathbf{E}_{\mathcal{X}} \left[G\left(\mathcal{X},n_\mathrm{r}\right)\right]$.
The reservation-only scheme is irrelevant to the on-demand price $c_\mathrm{s}$.
The optimal reservation, denoted by $n_\mathrm{ro}^*$, is then given by
\begin{equation}
n_\mathrm{ro}^* = \nabla U^{-1}\left(\frac{c_\mathrm{r}}{u_\mathrm{g} \mathbf{E}_{\mathcal{X}}\left[\Theta(\mathcal{X})\right]} \right).
\label{bench:1}
\end{equation}
The average surplus can be computed correspondingly.
In the on-demand-only scheme as used in \cite{2016:Nguyen}, the MVNO purchases spectrum dynamically according to the on-demand users and their locations.
The on-demand-only problem is a special case of the optimization problem (\ref{op:short}) with $n_\mathrm{r}=0$.
The optimal on-demand request, denoted by $n_\mathrm{so}^*$, is given by 
\begin{equation}
n_\mathrm{so}^* =\nabla U^{-1}\left(\frac{c_\mathrm{s}}{u_\mathrm{g} \Theta(\mathcal{X})} \right).
\label{bench:2}
\end{equation}
The surplus averaged over the period can be calculated correspondingly.

\begin{figure*}
	\centering
	\begin{subfigure}[b]{0.65\textwidth}
		\subcaption{}
		\includegraphics[width=\textwidth]{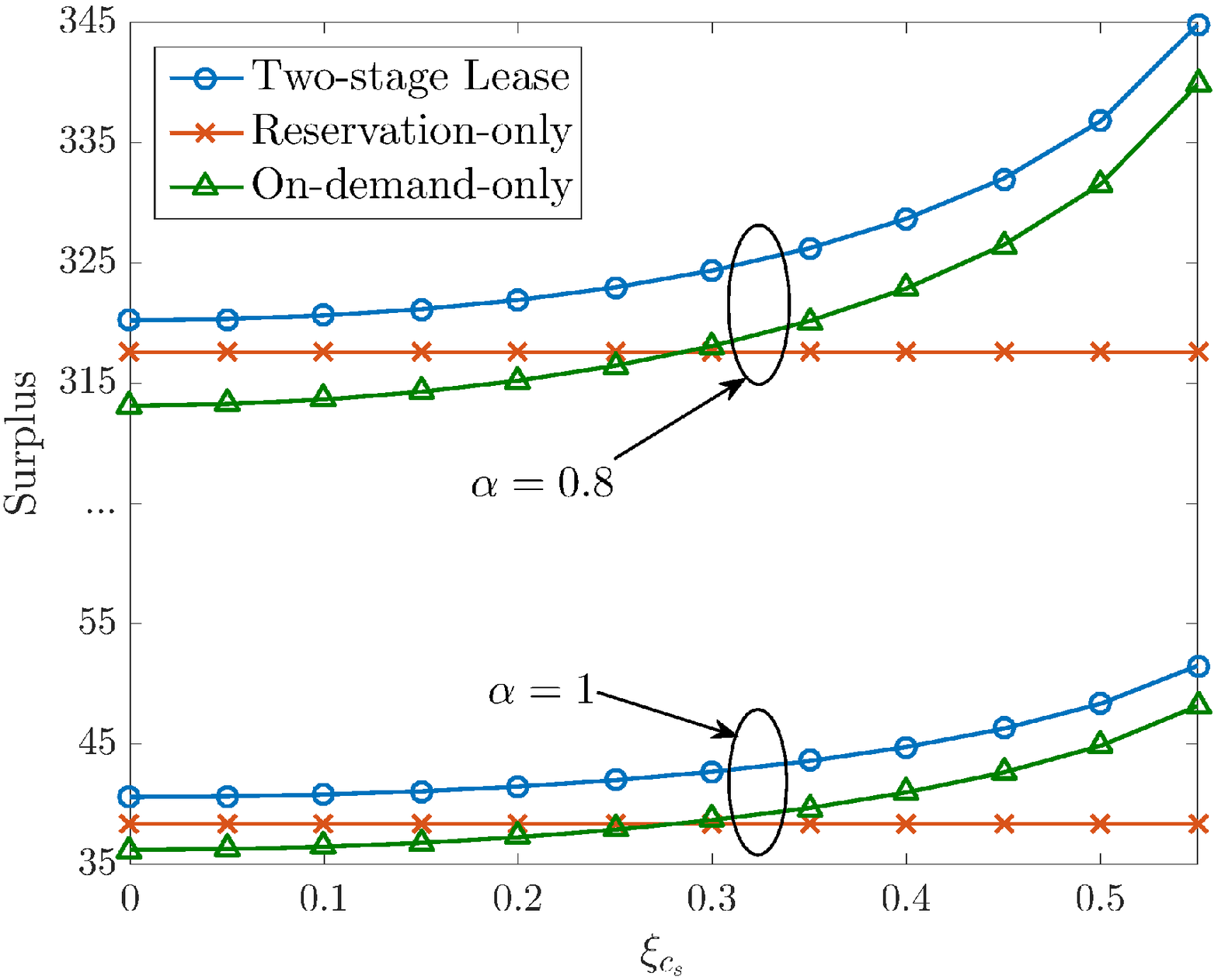}
		\label{fig:8a}
	\end{subfigure}
	\begin{subfigure}[b]{0.65\textwidth}
		\subcaption{}
		\includegraphics[width=\textwidth]{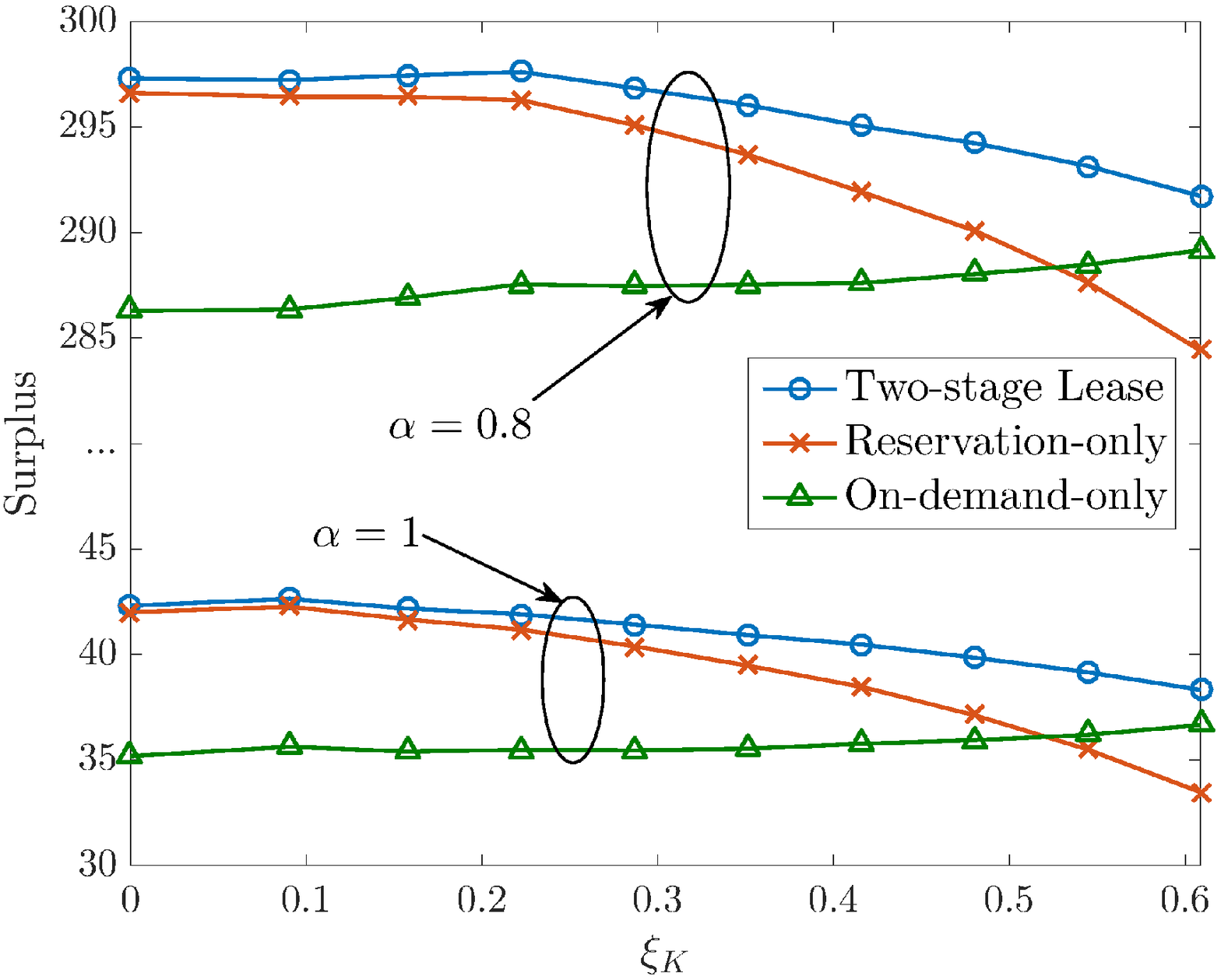}
		\label{fig:8b}
	\end{subfigure}
	\caption{\cmmt{Comparison of the two-stage leasing scheme with one-stage leasing schemes.
	(a) Surplus versus on-demand price variation $\xi_{c_\mathrm{s}}$. 
	(b) Surplus versus traffic variation $\xi_{K}$.}
 	}
	\label{fig:8}
\end{figure*}

\cmmt{
	We compare the surpluses achieved by the proposed two-stage leasing scheme with those by one-stage leasing schemes under different variation levels of on-demand price and user traffic in Fig. \ref{fig:8a} and \ref{fig:8b}, respectively.
	Since the reservation-only scheme is irrelevant to the on-demand price,
	the corresponding line remains constant for different $\xi_{c_\mathrm{s}}$ in Fig. \ref{fig:8a}.
	From both figures, we can see that the two-stage leasing scheme achieves much higher surplus than that of reservation-only scheme for all $\xi_{c_\mathrm{s}}$ and $\xi_{K}$.
	The differences becomes more apparent as $\xi_{c_\mathrm{s}}$ or $\xi_{K}$ increases.
	This is because that the on-demand stage preserves flexibility to acquire additional SCs to deal with traffic fluctuation.
	In this way, two-stage leasing scheme can reduce overbooking in the first stage.
	Moreover, we can see that the two-stage leasing scheme also outperforms on-demand-only scheme for all $\xi_{c_\mathrm{s}}$ and $\xi_{K}$. 
	The advantage is more distinct as $\xi_{c_\mathrm{s}}$ or $\xi_{K}$ becomes smaller.
	This is because that the first stage enables the MVNO to reserve SCs at a low cost.
	With the reserved SCs to serve a baseline amount of traffic, the MVNO can avoid purchasing expensive  on-demand SCs in real time.
	In sum, the two-stage leasing scheme can take advantage of both the low cost of advance reservation and the flexibility of on-demand request.
}

\section{Conclusions}
\label{sec:conclusion}
\cmmt{
	In this paper, we derived a two-stage leasing scheme, which enables the MVNO to take advantage of both the low cost of advance reservation and the flexibility of on-demand request.
	To find the optimal leasing decisions in the two stages,
	we formulated the problem as a tri-level nested optimization problem,
	and solved the problem efficiently.
	With our analysis, the MVNO can easily calculate the optimal amount of SCs to lease in each stage.
	Numerical results showed that the derived two-stage spectrum leasing scheme adapts to different levels of network variations, and achieves more surplus than conventional one-stage leasing schemes.
}

%\bibliographystyle{IEEEtran}
%\bibliography{../mvno}

\end{document}